\documentstyle{aipproc}
\include{psfig}

\begin{document}
\setcounter{page}{1}    
\pagenumbering{arabic}
\hspace*{10.7cm} DESY 97--260

\title{b-Quark Physics at DORIS \thanks{Talk presented at the $20^{th}$ 
Anniversary Symposium: Twenty Beautiful Years of Bottom Physics,
Chicago, June 29--July 2, 1997}}

\author{Dietrich Wegener}
\address{ Institut of Physics, University of Dortmund}
\maketitle

\section*{$\Upsilon$ -Physics - the early years 1977 - 1980}
The high energy physics program ($E_{cms} \le 8.6 GeV$) at DORIS was initiated by the PLUTO
collaboration which sent its proposal to the 
Forschungskollegium June 30, 1977 \cite {FK1}. The same day the observation of the 
$\Upsilon(9.46)$ resonance was announced to the public in a seminar at FNAL \cite{FNAL1}. 
The physics program proposed by PLUTO included the measurement of $\sigma_{tot}$ and the 
search for charm and $\tau$.  
The search for a $3^{rd}$--generation quark is not mentioned in the proposal. \\ 

The news from FNAL spread fast. The first documented discussion at DESY 
between machine physicists
and members of the PLUTO collaboration took place July 6, 1977 \cite{PROTO}. The energy upgrade 
of DORIS to $E_{cm} = 10$ GeV at moderate cost within half a year turned out to be possible, 
if parts of the new PETRA cavities and power supplies were used. Moreover, minor 
changes of the DORIS lattice were envisaged to avoid strong saturation effects of the magnets.
The proposal and its update \cite{UPDATE} were discussed by the DESY Forschungskollegium at
its meeting on July 15. The interest of measurements in the region of the new resonances was
emphasized and the directorate was urged to consider an upgrade of DORIS to $E_{cm} = 10 $ GeV
\cite{FK2}. Note in this context that PETRA was under construction at this time and
was scheduled to start running late summer 1978. \\

The possible physics program at a 10 GeV machine was discussed at a DESY workshop in 
October 1977. \cite {WORKS1}. J. B\"urger and H. Schr\"oder presented the physics 
program of the PLUTO and DASP~II collaboration, the latter just started 
to form. The ``Physics Priorities at DORIS'' from the theorists' point of view were 
discussed  by T. Walsh. Astonishingly  enough from todays point of view mainly the physics of 
the $2^{nd}$ generation was considered, only the $\Upsilon \rightarrow 3$--gluon decay
was briefly mentioned. Both experimental groups, on the other hand, discussed in detail
the possibility of learning of a $3^{rd}$--generation quark's properties in a
few days of running. \\

\begin{figure}
\centerline{\hbox{
  \psfig{figure=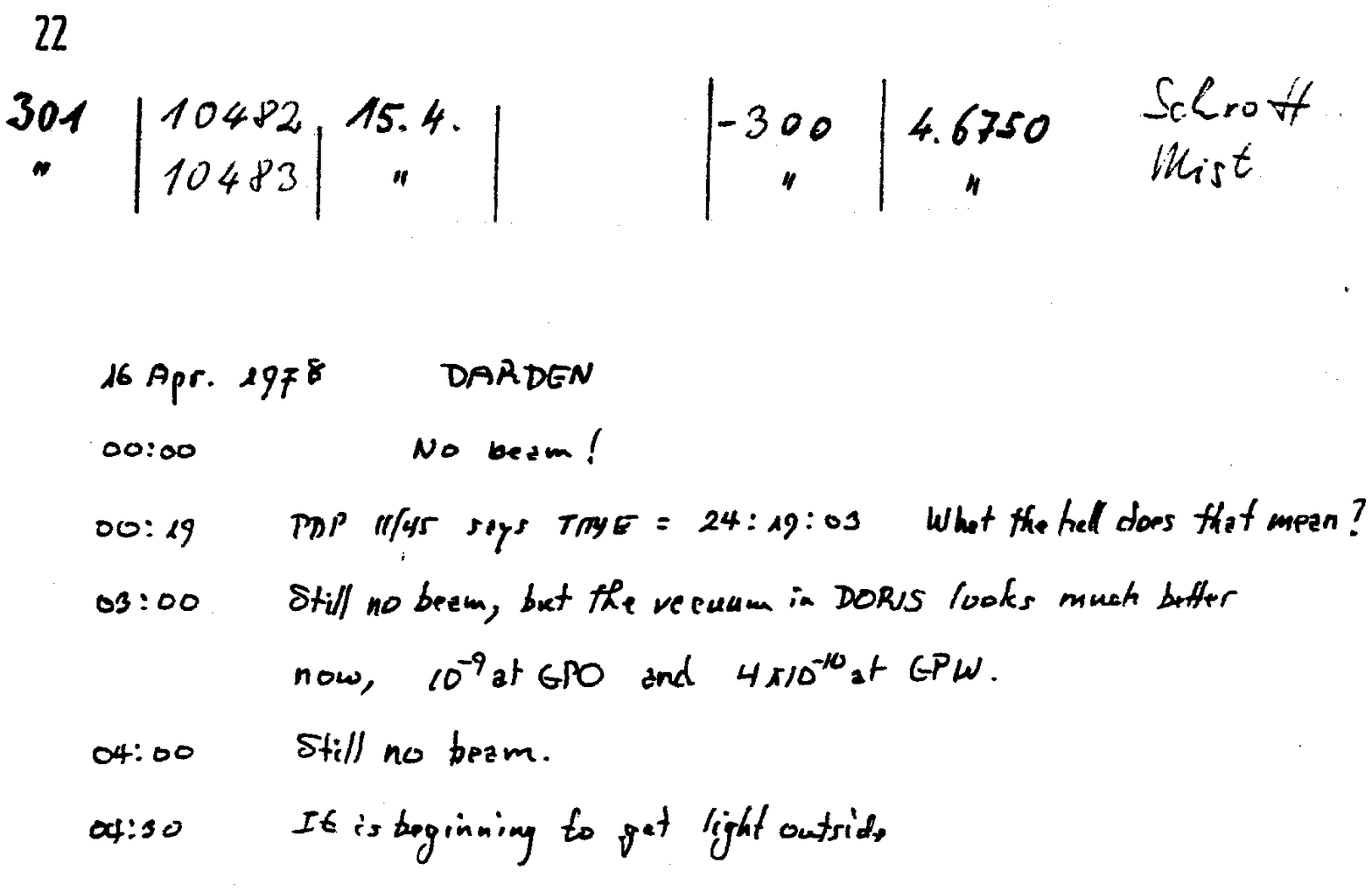,width=12cm}
}}
\caption{Copy from DASP~II runbook (15.4.1977)\em}
\label{fig1}
\end{figure}

The steps leading from the 5 GeV double--ring DORIS to the 10 GeV single--ring DORIS~I
are collected in table 1. 
The fast energy upgrade of DORIS was unexpected, I remember a seminar 
given by A. de Rujula at CERN in March 1978, where he discussed $\Upsilon$ physics. 
According to him the first experimental results were to be 
expected from CLEO early in 1979. The scan in the $\Upsilon(1S)$ region started at DESY 
April 15, 1978. Both
the machine and the detectors had problems in the beginning (fig.1).
A fluctuation observed first by DASP~II, and less prominently by the 
PLUTO collaboration after applying sophisticated cuts was convincing enough to motivate the 
DESY director to expend the first bottle of champagne. After a few days of running the
peak vanished, its trace can still be found in the smaller step size of the scan
around $9.38 $ GeV \cite {PLUTO1} in the published resonance curve. But finally, on April 30, 
the 
resonance signal was established. Why the Booze Up was delayed by 2 weeks (fig.2) cannot
be reconstructed any more. \\
\begin{figure}
\centerline{\hbox{
  \psfig{figure=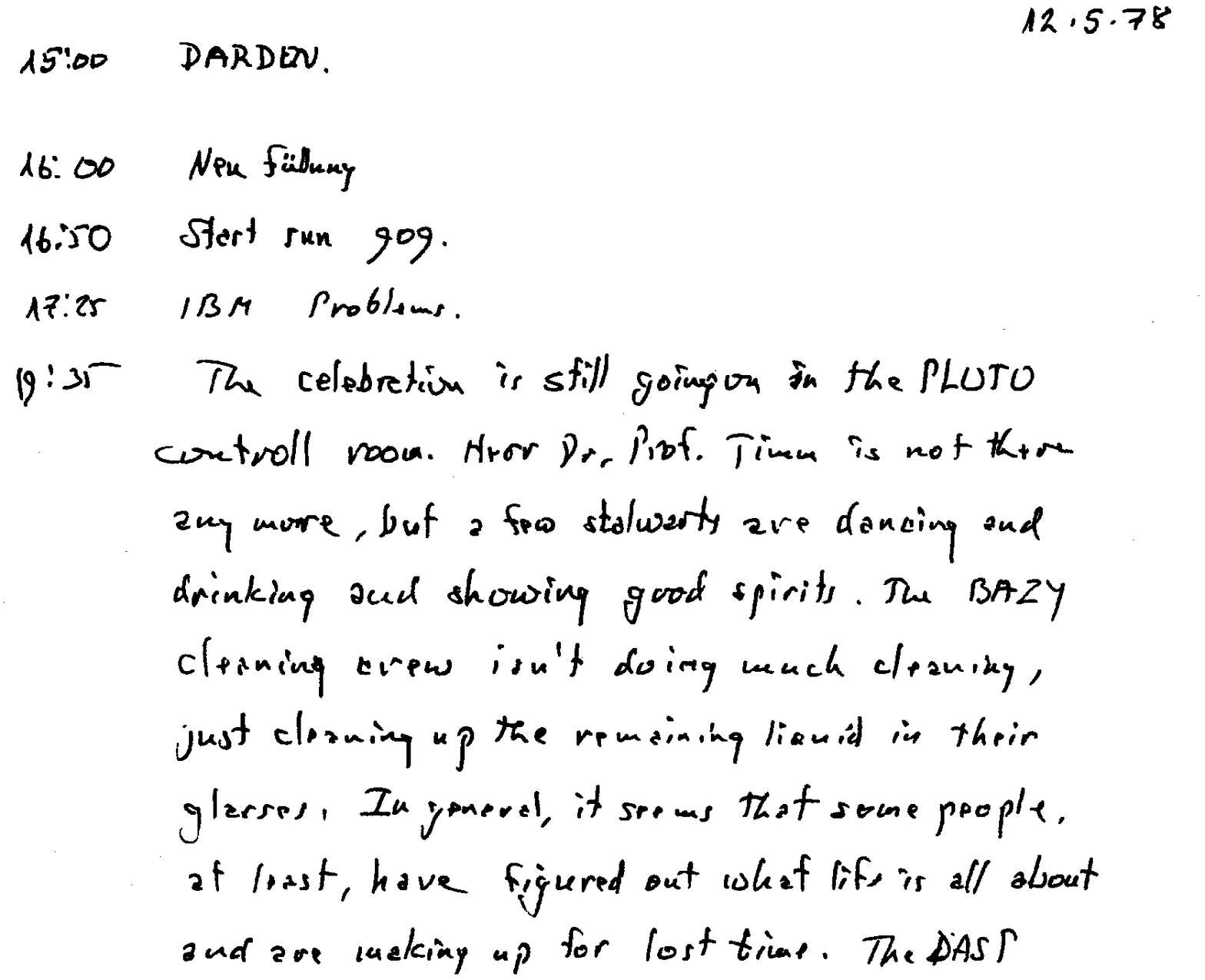,width=12cm}
}}
\caption{Copy from DASP~II runbook \em}
\label{fig2}
\end{figure}

The results proved that the resonance was narrow $\Gamma = (1.3 \pm 0.4)$ keV  
\cite{PLUTO1,DASP1}
and compatible with a $Q = - \frac{1}{3}$ charged quark. The mass of the resonant state was 
measured with high precision $M(\Upsilon(1S) = (9.46 \pm 0.01)$ GeV. These results 
established the 
$\Upsilon(9.46) $ resonance observed at FNAL \cite{FNAL2} as a $3^{rd}$--generation quarkonium 
state. A few months later the DASP~II and the LENA collaboration -- the latter replacing the 
PLUTO collaboration -- with marginal statistics determined the parameters of the 
$\Upsilon(2S)$ state \cite{DASP2,DESYH}. \\

After establishment of the quarkonium nature of the new resonances, the detailed study of the
hadronic decays was of special interest, since the resonance was predicted to decay mainly via
a 3--gluon final state \cite{DW1}. Already the first study of the event topology by PLUTO
revealed ``a striking change in mean sphericity and thrust on the 
$\Upsilon(9.46)$ resonance" \cite{PLUTO2}.  
The PLUTO collaboration addressed the problem of the 3--gluon final states in two further
papers \cite{PLUTO3}. In the first paper, received by the publisher in December 1978,
the authors concluded:
\begin{itemize}
\item The data are inconsistent with $\Upsilon$ decays into 2 light quarks (2--jet structure) 
      and into multipion phase space. 
\item All quantities related to momentum phase--space configurations are found to be
      in agreement with the proposed 3--gluon decay mechanism.
      Vector gluons are consistent with the proposed 3--gluon decays but not proven.
\end{itemize}
Summarizing, one might say that vector gluons as the field quanta of strong interaction
were not {\it discovered} at DORIS~I, but strong evidence for the decay of the 
$\Upsilon(1S)$ meson into three 
vector gluons {\it was} found \cite{DW2}. This point is missed in some papers describing 
the discovery of the gluon \cite{JARLS}. \\

The Crystal Ball (CB) \cite{BIENLEIN} and the ARGUS collaboration \cite{ARGUS1} later 
contributed further to our understanding of the $\mid b \bar{b} \rangle $ system. 

\section*{DORIS~II and its Detectors}
The major steps leading to the decision to upgrade DORIS~I and to increase the 
machine energy to 11.2 GeV (DORIS~II) are collected in table 1. 
The driving force was the growing interest in $B$ physics
and the possibility to upgrade DORIS~I at moderate cost and manpower \cite{WILLE1}.
An essential criterion for the final choice of the DORIS~II parameters was the 
requirement that the 
layout of the synchrotron--radiation beamlines was undisturbed. The essential changes of
DORIS~II with respect to DORIS~I were the decrease of the gap width and the increase 
of the number of coil windings of the magnets, 
thus reducing saturation effects and power consumption. The injection
was improved by installing separator plates and a faster kicker magnet. 
A major increase of the luminosity was achieved 
by mounting a special strong--focussing quadrupole at a small distance
from the interaction point \cite{WILLE2}.

\begin{table}
\caption{ DORIS storage ring }
\label{table1}
\begin{tabular}{ll}
6.7.1977      & First discussion to upgrade DORIS to $ 2 * 5$ GeV (DORIS~I) \\
              & Participants : Degele, B\"urger, Criegee, Fl\"ugge  \\
15.7.1977     & Forschungskollegium: strong support for upgrade  \\
16.12.1977    & Proposal to upgrade DORIS to $ 2 * 5$ GeV accepted \\
20.2.1978     &  Upgrade of DORIS starts $\rightarrow$ DORIS~I  \\
15.4.1978     & Scan in $\Upsilon(1S)$ region starts \\
30.4.1978     & $ \Upsilon(1S)$ resonance observed  \\
August 1978   & $ \Upsilon(2S)$ observed \\
July 1979     & Low--beta insertion to increase luminosity proposed by K. Wille \\
March 1980    & DORIS~I stops running for high energy physics  \\
February 1981 & DORIS~II ( 11.2 GeV machine) proposed \\
November 1981 & DORIS~II upgrade started  \\
May 1982      & DORIS~II starts running \\
1991          & DORIS~II by--pass upgrade for synchrotron radiation \\
October 1992  & ARGUS stops data taking \\
May 1993      & Tests to increase DORIS~II luminosity fail \\
              & high energy physics program at DORIS~II ends   \\
\end{tabular}
\end{table} 
With these improvements DORIS~II  achieved a maximal integrated luminosity of 
$1.8~ pb^{-1}$/day and an average luminosity of $0.5~pb^{-1}$/day. \\

The idea to build the ARGUS detector dates back to a dinner on September 14, 1977
\cite{DW3}. Already at the
DORIS workshop, one month later, the concept of "A New Detector at DORIS" including most  
of the features of the later ARGUS detector, was presented by 
W. Schmidt--Parzefall \cite{WORKS1}: 
\begin{itemize}
\item full coverage of the solid angle ($96$~\%)
\item good particle identification based on time--of--flight  and $dE/dx$ measurements
\item shower counters inside the solenoidal coil to detect photons of low energy $E_{\gamma}$
      $\ge 50$ MeV
\item $\mu$ chambers to detect muons with a momentum $p \ge 0.9$ GeV/c.
\end{itemize}

The ARGUS \footnote{The official interpretation is 
{\bf A}--{\bf R}ussian--{\bf G}erman--{\bf U}S--{\bf S}wedish 
collaboration, indicating the nationalities of the original proponants of the experiment. 
The unofficial 
interpretation by  one of the spouses knowing the senior members of the group too well
reads {\bf A}lle {\bf R}ichtigen {\bf G}enies {\bf U}nter {\bf S}ich} proposal 
was presented to the 
Forschungskollegium in October 1978 and accepted in July 1979.
The final design followed in many details the original idea, with only the layout of
the drift chamber improved to account for the requirements of optimal
pattern recognition. The physics benchmarks in the proposal were charm and $\tau$ physics. 
A detailed evaluation of a possible $B$ physics program was presented 
in April 1980 \cite{SCHR1}.
An expanded analysis of the possibilities  of studying $B$ physics with ARGUS 
followed in February 1981 \cite{Schuberth} when it
became clear that DORIS~I could be upgraded to an energy of 11.2 GeV.
The detector worked in a stable manner from 1982 through 1992. 
\begin{table}
\caption{ARGUS detector}
\label{table2}
\begin{tabular}{ll}
14.9.1977      & First plans to build ARGUS  \\
10.10.1977     & Meeting of DORIS Experiments  \\
               & Detector design study presented   \\
October 1978   & DESY proposal \#146 : ARGUS -- a new detector for DESY                \\
July 1979      & ARGUS proposal accepted by DESY directorate   \\
April 1980     & Interest in running ARGUS at 11.2 GeV emphasized \\
February 1981  & $B$ physics program at DORIS~II discussed  \\
6.10.1982      & ARGUS starts running    \\
September 1987 & $B^0 \bar{B^0}$ mixing observed  \\
Autumn 1989    & Observation of $ b \rightarrow u$ transitions \\
8.10.1992      & ARGUS stops data taking   \\
\end{tabular}
\end{table}               
\\

During the DORIS workshop in February 1981 the idea arose to transfer the Crystal Ball~(CB) 
detector from SLAC to DESY \cite{BIENLEIN}. The proposal was soon presented and accepted 
in summer 1981. The CB detector was transported to DESY in spring 1982 and started data 
taking on August 6, 1982, while ARGUS rolled in two months later. 
The competition between the two 
experiments delayed the $B$ physics program at DORIS for nearly 3 years because the 
CB collaboration preferred to run at the energy of the $\Upsilon(1S)$ and 
$\Upsilon(2S)$ resonance,
since its detector was optimized for spectroscopic studies. As shown by Table 3 in the first 
\begin{table}
\caption{Integrated luminosity collected 1983/1986 at DORIS~II}
\label{table3}
\begin{tabular} {lcccc}
\hline
 &1983 & 1984 & 1985 & 1986  \\[0.4ex]
\hline
$\Upsilon(1S)$ & ~9 pb~-1 & 23 pb~-1 & - & 31 pb~-1 \\ [0.4ex]
\hline
$\Upsilon(2S)$ & 27 pb~-1 & 25 pb~-1 & - & -     \\[0.4ex] 
\hline 
$ \Upsilon(4S)$ &  ~6 pb~-1 & 14 pb~-1 & 45 pb~-1 & 44 pb~-1 \\ [0.4ex]
\hline
Continuum & ~4 pb~-1 & ~7 pb~-1 & 16 pb~-1 & 19 pb~-1 \\ [0.4ex]
\hline
\end{tabular}
\end{table}
years of DORIS~II running, priority was given to the CB physics program.  
The following facts may have contributed to the decision:
\begin{itemize}
\item CB was a running detector with a respectable record of discoveries.
\item It was an established and successful collaboration while the ARGUS senior 
      members at that time were youngsters.
\item CB observed an unexpected signal \cite{ZETA} and hopes were running high for a short
      time that a light Higgs had been discovered 
\footnote{At this place it is appropriate to remind the reader of 
the guidelines for searches formulated 200 years ago: ``one may notice that a shrewd
intellect brings more artifice to bear the fewer data are available; indeed, to demonstrate
his mastery he will select from all available data only those few favorable to his views;
the remainder he will arrange so as not to obviously contradict his conclusions; and
finally hostile data will be isolated, surrounded and disarmed'' \cite{GOETHE}.}.   
Unfortunately, the result turned out to be irreproducible \cite{DESYann}.
\end{itemize}
Before discussing the most important ARGUS discovery a further obstacle met by ARGUS should be
mentioned. As shown in fig.3 two major gaps in the data taking are manifest. They 
follow the most important ARGUS discoveries: 1987 $B^0 \bar{B^0}$ mixing was observed, 1989
$b \rightarrow u $ transitions were detected. One might wonder if the DESY directorate 
suspected ARGUS was not putting enough emphasis on analysis, and therefore wanted to give 
the collaboration a 
chance to improve in this respect. Note, however, that the official explanation is different: 
1987 HERA got priority and 1990/1991 the DORIS bypass was built. From the latter 
``improvement'' the machine never recovered for high energy running.

\begin{figure}
\centerline{\hbox{
  \psfig{figure=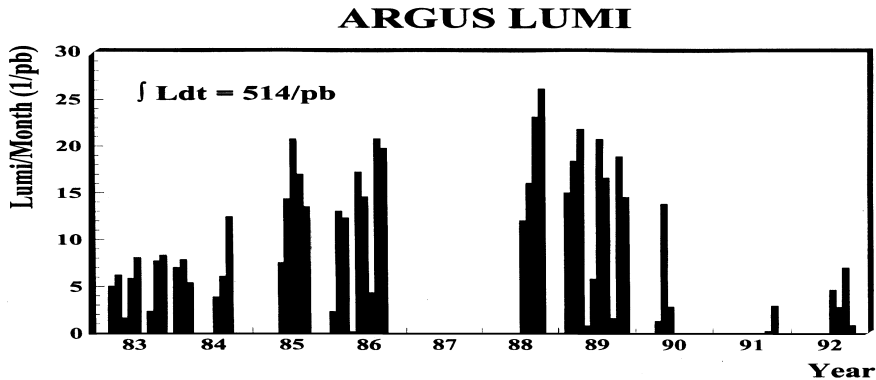,height=5cm,width=12cm}
}}
\caption{Luminosity collected by ARGUS 1982--1992 \em}
\label{fig3}
\end{figure}

\section*{Discoveries}
The ARGUS collaboration for more than one decade substantially contributed to different fields 
of high energy physics. The results are summarized in \cite{ARGUS1}. In $B$ physics the
highlights are the following ``firsts":
\begin{itemize}
\item  Observation of $B^0 \bar{B^0}$ mixing \cite{ARGUS2}
\item  Observation of charmless $B$ decays \cite{ARGUS3}
\item  Reconstruction of exclusive semileptonic $B$ decays to $D^*$ and $D$ mesons 
       respectively \cite{ARGUS4}
\item  Reconstruction of exclusive hadronic $B$ decays \cite{ARGUS41}
\item  Model--independent measurement of semileptonic $B$ decays \cite{ARGUS5}
\item  Observation of charmed baryons in $B$ decays \cite{ARGUS6}
\end{itemize}
Due to a lack of time only the most important discovery is discussed in some detail.

\subsection*{$B^0 \bar{B^0}$ Mixing} 
Present universal interest in $B$ physics is largely due to the discovery of 
$B^0 \bar{B^0}$ mixing
by the ARGUS collaboration. As is well known \cite{SCHR2} the process is mediated by box 
diagrams. The mixing parameter $r_d$ derived from time integrated measurements is given by
the expression
\begin{equation}
r_d = \frac{N(B^0 \rightarrow \bar{B^0})}{N(B^0 \rightarrow B^0)} = \frac{(\Delta M \tau_B)^2}
{(2 + \Delta M \tau_B)^2} \sim m_{top}^4 
\end{equation}
\begin{equation}
\Delta M = \frac{G_F^2}{6 \pi^2}~ B_B f_B^2 m_B \mid V^*_{tb} V_{td} \mid^2 m_{top}^2 
F~\left( \frac{m_{top}^2}{m_W^2} \right)~ \eta_{QCD}
\end{equation}
i.e. mixing is dominated by virtual t--quark exchange. The experimental situation in 1986  
was as follows:
PETRA experiments did not observe a signal, i.e. $m_{top}\le 23.3$ GeV, 
while UA1 claimed \cite{UA1} a signal at  $m_{top} \approx 40$ GeV.
As a consequence a small mixing parameter $r_d \approx 0.01$ was expected. A scan of the 
literature by the 
author in September 1985, while preparing a memo to the DESY PRC, showed that under 
optimistic assumptions on $f_B$ a mixing of $r_d \le 0.05$ was predicted \cite{ARGUS7}. 
Mixing searches using b--quark jets by MARK II, MAC and UA1 were not conclusive. \\

In summer 1986, for the first time ARGUS and CLEO 
had enough statistics to exploit the 
particularly clean conditions at the $\Upsilon(4S)$ to search for $B^0 \bar{B^0}$ mixing.
The semileptonic decay $B^0 (\bar{b}d) \rightarrow l^+ X$ served as tag of the heavy 
flavor, i.e. $l^{\pm}l^{\pm}$ and $l^+l^-$ events were used to measure the mixed and 
unmixed events respectively. At the Berkeley conference the groups presented their limits  
(90~\% CL): 
$r_d \le 0.12 $ (ARGUS \cite{ARGUS8}) and
$r_d \le 0.20$ (CLEO\cite{CLEO1}).
Immediately after the conference ARGUS prepared a publication which even got a DESY number 
(DESY 86--121). However, the distribution of the paper was stopped at the last moment by 
H. Schr\"oder. He collected all preprints at the moment they left the 
printer's office. All copies were burnt! \\
 
What observation led to this reaction? In August 1986 H. Schr\"oder started an analysis of the 
$\bar{B^0} \rightarrow D^{*+} l^- \bar{\nu_l}$ decay, which was of special interest, since a 
large branching ratio of $\sim 8$~\% was predicted but no measurements existed. 
Since the $D^{*+}$ reconstruction capabilities of ARGUS were excellent and $e$ and $\mu$ 
were identified with high
efficiency, a high--
statistics $\bar{B^0} \rightarrow D^{*+} l^- \bar{\nu_l}$ sample out
of $\sim 25000$ $B^0 \bar{B^0}$ events was expected. However, a new method had to
be developed to reconstruct 
these events with an undetected $\nu_l$, whose mass can be derived from the measurements:

\begin{equation}
m_{\nu}^2 = (E_B -E_{D^{*+}} -E_{l^-})^2 -(\vec{p_B} - (\vec{P_{D^{*+}}}+\vec{P_{l^-}}))^2
\end{equation}
From the first successful reconstruction of exclusive hadronic $B$ decays \cite{ARGUS41} 
it was known that
\begin{equation}
2 E_B = m(\Upsilon(4S)) \approx 2 m_B\, .
\end{equation}
Since $E_B =E_{beam}$, $\mid\vec{p_B} \mid = 0.33$ GeV/c and hence can be neglected in (3). 
Therefore,
\begin{equation}
m^2_{\nu} \approx M^2_{rec} = (E_{beam} -E_{D^{*+}}-E_{l^-})^2 -
(\vec{p_{D^{*+}}}+\vec{p_{l^-}})^2
\end{equation}
\begin{figure}[h]   
\centerline{\hbox{
  \psfig{figure=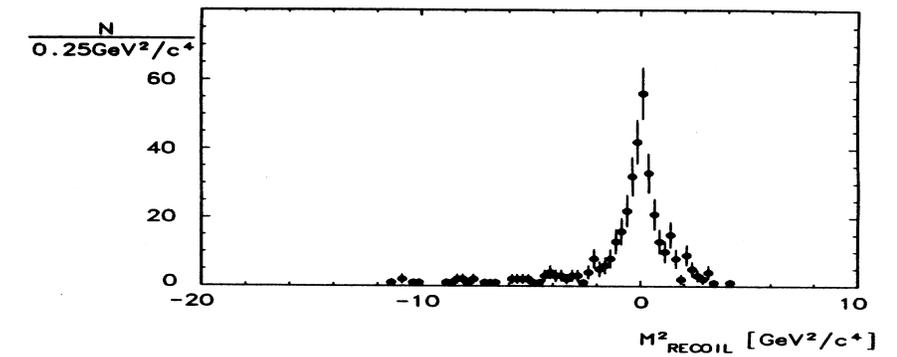,height=5cm,width=12cm}  
}}
\caption{Measured recoil mass distribution \em}
\label{fig4}
\end{figure}

As expected, a peak at $M_{rec}^2 \approx 0$ is observed with small, wel--known 
background (fig.4).
Though the application of this method was controversial \cite{STONE}, in the following
years it was applied in many analyses by the CLEO and the ARGUS collaboration. In September 
1986, 50 events with a reconstructed $B^0(\bar{B^0})$ 
were available to tag the heavy flavor of the
$B^0$. H. Schr\"oder presented the first results of his analysis at the ARGUS group meeting
on September 25, 1986 (fig.5). He studied in detail the events with a full reconstructed 
$B^0(\bar{B^0})$ meson. The observed multiplicity, number of kaons and leptons followed
the expectation, but he stumbled over a few events with wrong charged kaons and leptons
respectively. In the data sample five candidates for mixed events were observed: 
$2 B^0e^+$, 2$ \bar{B^0}e^-$, $1\bar{B^0} \mu^-$
besides 23 candidates for unmixed events. After background subtraction 
a mixing ratio of $r_d = 0.20 \pm 0.12$ was obtained. 

\begin{figure}[h]
\centerline{\hbox{
  \psfig{figure=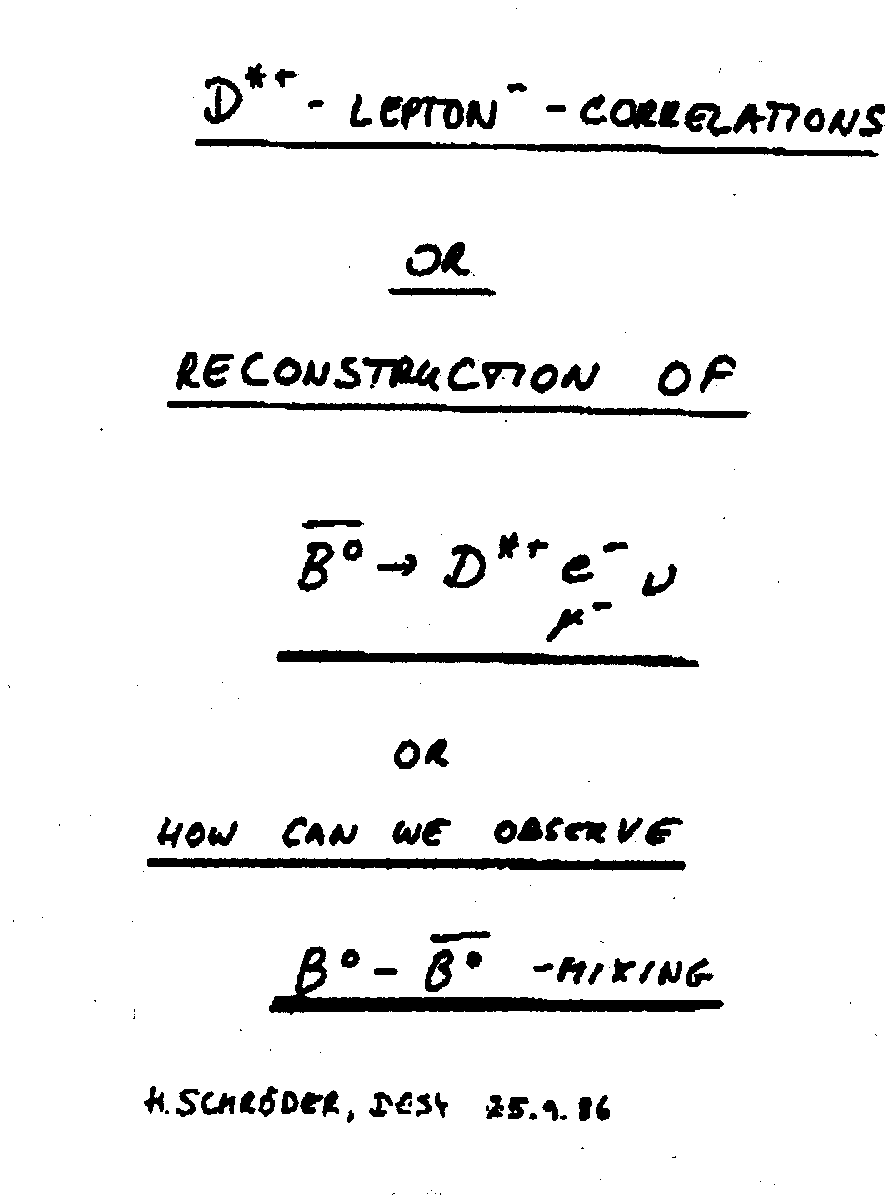,height=6cm}  
}}
\caption{First page of H. Schr\"oder's talk announcing the observation of 
$B^0 \bar{B^0}$ mixing \em }
\label{fig5}
\end{figure}

The claim that $B^0 \bar{B^0}$ mixing had indeed been observed was supported by the 
observation of one full reconstructed event with 2 $\bar{B^0}$ mesons in the final 
state decaying via $\bar{B^0} \rightarrow D^* \mu^+ \bar{\nu}_{\mu}$ (fig.6). 
Both $\mu^+$ and the $K^+$ meson were uniquely identified. 
The observation of this event is a convincing example of the 
advantages of the ARGUS detector: precise momentum measurement, good particle identification
and hermiticity. The observation of $D^*$--lepton correlation therefore provided an extremely 
useful tool. This proved the existence of $B^0 \bar{B^0}$ mixing with a large mixing parameter,
totally unexpected at that time. 

\begin{figure}[h]
\centerline{\hbox{
  \psfig{figure=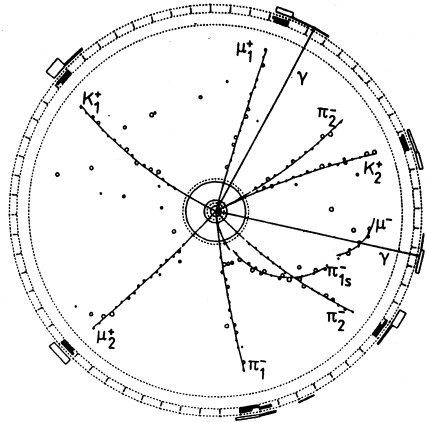,width=6cm}  
}}
\caption{First full reconstructed $B^0 \bar{B^0}$ mixing event \em}
\label{fig6}
\end{figure}

This result stimulated further activity. Y. Zaitsev and A. Golutvin repeated the same--sign 
lepton pair analysis. A signal was observed in this sample as well. The major 
improvement compared
to the previous analysis presented at Berkeley \cite{ARGUS8} was the increase in the 
collected luminosity of more than a factor of 2. 
Furthermore, the better understanding of the detector allowed improving the cuts applied
in the analysis. The mixing parameter derived in this analysis was in good agreement with
the result of the exclusive analysis. Combining the results ARGUS got
\begin{equation}
r_d = (0.21 \pm 0.08)
\end{equation} 
in good agreement with the present world average \cite{PDG96}.
\\

To explain the large mixing parameter, ARGUS had to assume the top mass to be large, 
$m_{top} > 50$ GeV, 10 years ago an unconventional assumption in view of the UA1 claim
\cite{UA1}.
The paper was published on June 25, 1987, just 10 years after the discovery of the 
$\Upsilon(1S)$ resonance by Lederman and coworkers at FNAL. The large
mixing in the $B$ system raised hopes of observing CP violation in this system,
a prospect attracting many scientists to the field. The experiments 
presently under construction \cite{CPB}
underline the importance of the seminal ARGUS result obtained 10 years ago.

\section*{Summary}
I will abstain from discussing in detail the other important contributions of ARGUS to
$B$ physics, I only want to address the question why the collaboration was so successful 
for nearly 10 years. The answer was given by David Cassel in his talk ``The Impact
of ARGUS on Experimental Heavy Flavor Physics" \cite{CASSEL}, where he discussed the lessons
to be learned from ARGUS: 
\begin{itemize}
\item Have a better detector that can see ``all".
\item Learn to use the hermiticity of the detector.
\item Have excellent physics ideas and follow them.
\item Have excellent physics analysis software.
\item Have a little bit of luck.
\item Do not underestimate the competition.
\end{itemize}
There are a bit too many {\it excellent's} in this list but otherwise I have nothing to add. 
Hopefully the new generation of experiments will be as prolific
as the $2^{nd}$ generation, and the participants will have as much fun as the CLEO and ARGUS
collaborations had.

\subsection*{Acknowledgement}
I thank the organizers for their generous hospitality and for a stimulating conference,
where I learned many secrets of those experiments which established 
the field of $B$ physics. Thanks for
discussions, unpublished material, etc. to J. B\"urger, H. Meyer, W. Schmidt--Parzefall,
K. Wacker and K. Wille. Most helpful in reconstructing the steps leading to the 
discovery of $B^0 \bar{B^0}$ mixing was the information I got from H. Schr\"oder. 
This work was supported by the BMBF under contract number 6DO57I.



\end{document}